# Equation for the Abraham force in non-conducting medium and methods for its measurement

Yurii A. Spirichev

Research and Design Institute of Radio-Electronic Engineering - branch of Federal State Unitary Enterprise of Federal Scientific-Production Center "Production Association "Start" named after Michael V.Protsenko"

E-mail: yurii.spirichev@mail.ru

(Dated: April 7, 2017)

**Abstract**

This article is devoted to obtaining an equation for the Abraham force in a continuous non-conducting medium and methods of its measurement. The equation for the Abraham force is obtained from the Minkowski tensor. The Abraham force appears when the vectors D and E, H and B are non-collinear. From the equation for the Abraham force it follows that it is a vortex force, and its divergence is equal to zero. It shows the existence of the Abraham electric force and the magnetic Abraham's force. Various measurement methods of the Abraham force, which follow from its equation, are given.

**The contents**

    1. **Introduction**

    The problem of the interaction of electromagnetic fields with matter and the resulting forces has been discussed for many years, but there is no a unique opinion on this problem. In recent years, attempts to create materials with unique electromagnetic features are being undertaken. Therefore, the issue of interaction of electromagnetic field with medium has gone to the frontburner. Volumetric electromagnetic forces, especially the power of Abraham, cause a lot of disputes. Volumetric electromagnetic forces of interaction of electromagnetic field with substance are defined by electromagnetic field momentum in the medium. In general form, the Abraham force $\mathbf{F}_A$ is written as the difference of expressions for derivatives with respect to time of the electromagnetic momentum in

the Minkowski and Abraham forms $\mathbf{F}_A = \partial_t \mathbf{g}^M - \partial_t \mathbf{g}^A$, where $\mathbf{g}^M = (\mathbf{D} \times \mathbf{B})/4\pi \cdot c$ - the electromagnetic momentum in the Minkowski form, $\mathbf{g}^A = (\mathbf{E} \times \mathbf{H})/4\pi \cdot c$ - electromagnetic momentum in the Abraham form, **D**, **B** - vectors of electric and magnetic induction, and **E**, **H** – intensity of electric and magnetic fields. Many scientific studies, both theoretical and experimental, are dedicated to volumetric electromagnetic forces and the Abraham force [1-35]. It is believed that the Abraham force is measured and determined, but there are studies that challenge the results of the experiments. Moreover, there are studies that cast doubt on the very existence of the Abraham force in nature. Thus, opinions on the Abraham force differ and there is no consensus on this matter. The reason is the lack of proper equation for the Abraham force, which leads to different opinions and disputes on this matter, as well as to the incorrectness of the experimental work on its discovery.

This article is devoted to the solving of the Abraham force problem and obtaining of its equation from Minkowski energy-momentum tensor, as well as the methods of its measurement.

## 2. Minkowski energy-momentum tensor and its decomposition

The canonical energy-momentum tensor in the general form can be written as:

$$T_{\nu\mu} = \begin{bmatrix} W & i\frac{1}{c}\mathbf{S} \\ ic \cdot \mathbf{g} & t_{ik} \end{bmatrix} \quad (\nu, \mu = 0, 1, 2, 3; \; i, k = 1, 2, 3) \tag{1}$$

where  **W** – energy density;

**S** – the energy flux density (Poynting vector);

**g** – the momentum density;

$t_{ik}$ – density momentum flux tensor (the tension tensor).

Components of the energy-momentum tensor (1) in the form of Minkowski have the form:

$$W = (\mathbf{E} \cdot \mathbf{D} + \mathbf{H} \cdot \mathbf{B})/2 \qquad \mathbf{S} = \mathbf{E} \times \mathbf{H}$$

$$\mathbf{g}^M = \mathbf{D} \times \mathbf{B} \qquad t_{ik}^M = (E_i D_k + H_i B_k) - \delta_{ik}(\mathbf{E} \cdot \mathbf{D} + \mathbf{H} \cdot \mathbf{B})/2.$$

The Minkowski tensor is asymmetric and can be decomposed into antisymmetric and symmetric energy-momentum tensors:

$$T_{\nu\mu} = \frac{1}{2} \cdot T_{[\nu\mu]} + \frac{1}{2} \cdot T_{(\nu\mu)} = \frac{1}{2} \cdot \begin{bmatrix} 0 & i\frac{1}{c}\mathbf{S} - ic \cdot \mathbf{g} \\ -i\frac{1}{c}\mathbf{S} + ic \cdot \mathbf{g} & t_{ik} - t_{ki} \end{bmatrix} + \frac{1}{2} \cdot \begin{bmatrix} 2W & i\frac{1}{c}\mathbf{S} + ic \cdot \mathbf{g} \\ i\frac{1}{c}\mathbf{S} + ic \cdot \mathbf{g} & t_{ik} + t_{ki} \end{bmatrix} \tag{2}$$

Antisymmetric energy-momentum tensor $T_{[\nu\mu]}$ describes the energy of four-dimensional rotation of the medium, and the symmetric tensor $T_{(\nu\mu)}$ describes the energy of a four-dimensional deformation of the medium under the influence of the electromagnetic field.

## 3. The equation for the Abraham force in a non-conductive medium

Taking the four-dimensional divergence of the tensors (2), we obtain the equation for conservation of energy and momentum for four-dimensional rotation of the medium and the medium deformation under the influence of the electromagnetic field.

From the antisymmetric energy and momentum tensor $T_{[\nu\mu]}$ follows the equations of conservation of energy and electromagnetic momentum for four-dimensional rotation medium:

$$\nabla \cdot (\mathbf{g}^M - \frac{1}{c^2}\mathbf{S}) = 0 \quad \text{or} \quad \nabla \cdot (\mathbf{g}^M - \mathbf{g}^A) = 0 \tag{3}$$

$$\partial_t(\mathbf{g}^M - \frac{1}{c^2}\mathbf{S}) - \frac{1}{4\pi}\nabla \times (\mathbf{E} \times \mathbf{D} + \mathbf{B} \times \mathbf{H}) = 0 \quad \text{or} \quad \partial_t\mathbf{g}^M - \partial_t\mathbf{g}^A = \frac{1}{4\pi}\nabla \times (\mathbf{E} \times \mathbf{D} + \mathbf{B} \times \mathbf{H}) \tag{4}$$

Let's take time derivative of equation (3) and get the equation:

$$\nabla \cdot (\partial_t\mathbf{g}^M - \partial_t\mathbf{g}^A) = 0 \quad \text{или} \quad \nabla \cdot \mathbf{F}_A = 0 \tag{5}$$

The expression in brackets represents the Abraham force. From this equation it follows that the divergence of the Abraham force is equal to zero, i.e. this force has a vortex character.

Equation (4) is the equation for the Abraham force:

$$\mathbf{F}_A = \frac{1}{4\pi}\nabla \times (\mathbf{E} \times \mathbf{D} + \mathbf{B} \times \mathbf{H}) \tag{6}$$

Equation (6) confirms the conclusion which was drawn from equation (5) and means that the divergence of the Abraham force is equal to zero since it has a vortex character. From equation (6) it follows that the power of Abraham appears when the vectors $\mathbf{D}$ and $\mathbf{E}$, $\mathbf{H}$ and $\mathbf{B}$ are non-collinear. Equation (6) shows that the Abraham force consists of two vortex parts: electric and magnetic. Thus, it is possible to distinguish electric Abraham force $\mathbf{F}_A^E$ and magnetic Abraham force $\mathbf{F}_A^M$:

$$\mathbf{F}_A^E = \frac{1}{4\pi}\nabla \times (\mathbf{E} \times \mathbf{D}) \quad \text{and} \quad \mathbf{F}_A^M = \frac{1}{4\pi}\nabla \times (\mathbf{B} \times \mathbf{H}) \tag{7}$$

This distinction is of interest in the experiments on the Abraham force measurement. Depending on the electromagnetic characteristics of the medium, either electric or magnetic part of the Abraham force may be missing.

In equation (6) for the Abraham force, there are no restrictions on constitutive equations, and it is universal for any non-conductive continuous medium. Equation (6) implies that if a non-conductive medium is described by the canonical material equations of the form $\mathbf{D} = \varepsilon \cdot \varepsilon_0 \cdot \mathbf{E}$ and $\mathbf{H} = \mathbf{B}/\mu \cdot \mu_0$, and the relative dielectric and magnetic permeability of the medium $\varepsilon$ and $\mu$ are constant or scalar functions, than the vectors $\mathbf{D}$ and $\mathbf{E}$, $\mathbf{H}$ and $\mathbf{B}$ are collinear and the Abraham force is equal to zero.

From the symmetric tensor $T_{(\nu\mu)}$ follows the equation for conservation of electromagnetic energy and momentum and four-dimensional deformation of the medium under the influence of electromagnetic field:

$$2\frac{1}{c^2}\partial_t W - \nabla \cdot (\mathbf{g} + \frac{1}{c^2}\mathbf{S}) = 0 \quad \text{и} \quad \partial_t(\mathbf{g} + \frac{1}{c^2}\mathbf{S}) - \partial_i(t_{ik} + t_{ki}) = 0$$

In this article, these equations are not considered.

## 4. Measurement methods for the Abraham force in a non-conductive medium

From equation (6) it follows that for setting up experiments on the Abraham force measurement it is needed to choose a medium in which electromagnetic characteristics provide the rotation of the field and induction vectors relative to each other. Thus, the maximum magnitude of the Abraham force is provided when the field vector and induction vector are **turned** relative to each other at an angle $\pi/2$. Since the vector angular momentum, created by the vortex Abraham force, is directed along the vector of electromagnetic field momentum, the vectors of the excitation electric field **E** and magnetic induction **B** have to be orthogonal with respect to each other and the axis of the torsional pendulum of the measurement system.

From equations (6) and (7) follow different ways of measurement of the Abraham force. These methods can be divided according to the types of the Abraham force excitation in the medium by electromagnetic field:

a) the electric field **E** is alternating, and the magnetic induction **B** is constant;

b) magnetic induction **B** is variable, and the electric field **E** is a constant;

c) the electric field **E** and magnetic induction **B** are variables.

Methods of measurement may also be classified according to the electromagnetic characteristics of the non-conductive medium:

a) the electromagnetic characteristics of the medium provide rotation only of the **electric** displacement vector **D** relative to the electric field vector **E**;

b) the electromagnetic characteristics of the medium provide rotation only of the magnetic field vector **H** relative to the magnetic induction vector **B**;

c) the electromagnetic characteristics of the medium provide rotation of the electric displacement vector **D** relative to the electric field vector **E** and rotation of the magnetic field vector **H** relative to the magnetic induction vector **B**.

The specific way of Abraham force measurement is determined by the type of medium excitation and the electromagnetic characteristics of non-conductive medium. The alternating electromagnetic field used to excite the Abraham force should be submitted with the frequency of mechanical resonance of the measurement system. When using the measurement method that takes into account both electric and magnetic characteristics of the medium it is necessary to choose medium with the turn of vectors **D** and **H** in one and the same direction.

Optical method of Abraham force measurement is based on the use of laser radiation, which is injected from the end face into optically transparent cylindrical rod along its axis. Cylindrical rod suspended in a vertical position in the form of a torsional pendulum. The rod is made of a material that provides rotation of the field and induction vectors in relation to each other at an angle. When you turn on the laser radiation, the cylindrical rod will be influenced by the turning Abraham force. To enhance the effect, laser pulse is fed with the frequency of mechanical resonance of the suspended rod.

## 5. Conclusion

From the Minkowski energy-momentum tensor follows the equation for the Abraham force which appears in continuous non-conducting medium under the influence of alternating electromagnetic field. The Abraham force appears when the vectors **D** and **E**, **H** and **B** are non-collinear. Equation for Abraham force shows that it is a vortex force, and its divergence is equal to zero. This equation also shows the existence of two Abraham forces; the electric force and the magnetic force. Different ways of Abraham force measurement appearing from the equation are shown.